\documentclass[twocolumn,preprintnumbers,prl]{revtex4}
\usepackage{graphicx}
\usepackage[hypertex]{hyperref}

\begin{document}

\title{Gauge Mediation Simplified}

\author{Hitoshi Murayama and Yasunori Nomura}
\affiliation{Department of Physics, University of California,
                Berkeley, CA 94720, USA}
\affiliation{Theoretical Physics Group, Lawrence Berkeley National Laboratory,
                Berkeley, CA 94720, USA}

\begin{abstract}
  Gauge mediation of supersymmetry breaking is drastically simplified 
  using generic superpotentials without $U(1)_R$ symmetry by allowing 
  metastable vacua.
\end{abstract}
\maketitle

Breaking supersymmetry has been a non-trivial task.  A general argument 
by Nelson and Seiberg is that it requires a theory with a continuous 
exact $U(1)_R$ symmetry if we assume that the superpotential is 
generic~\cite{Nelson:1993nf}.  In addition, an argument based on 
the Witten index~\cite{Witten:1981nf} said that the theory must be 
chiral.  This is because one can continuously deform a vector-like 
theory by mass terms to a pure Yang-Mills theory, which is known to 
have a finite Witten index (dual Coxeter number) and hence supersymmetric 
vacua.  Chirality and $U(1)_R$ invariance strongly limit the choice 
of possible theories that break supersymmetry.  Therefore explicit 
models of supersymmetry breaking appear rather special and hence do 
not seem likely to come out from a more fundamental theory such as 
string theory.  This problem is exacerbated by the fact that the 
supersymmetry breaking sector should couple to the standard model 
multiplets to induce soft supersymmetry breaking parameters in 
a flavor-independent fashion.

Later, vector-like models were found~\cite{Izawa:1996pk}.  They evade 
the Witten index argument because the mass terms can always be absorbed 
by shifting singlet fields in the theory.  The required superpotential, 
however, is not generic unless one imposes an exact $U(1)_R$ symmetry.

The requirement of an exact $U(1)_R$ symmetry is unfortunate, because 
exact global symmetries are not expected to exist in quantum theory of 
gravity such as the field-theory limit of string theory.  In addition, 
embedding a model of supersymmetry breaking into supergravity requires 
explicit breaking of $U(1)_R$ to allow for a constant term in the 
superpotential needed for canceling the cosmological constant.  Once 
$U(1)_R$ is not an exact symmetry, it is not clear how one can justify 
the form of the superpotential required for supersymmetry breaking.

In this letter, we advocate to discard $U(1)_R$ symmetry altogether 
from the theory, and allow for completely generic superpotentials. 
According to the Nelson--Seiberg argument, such a theory would not 
break supersymmetry.  Yet, it may have a {\it local}\/ supersymmetry 
breaking minimum.  Supersymmetry is broken if the low-energy limit of 
the supersymmetry breaking sector has an {\it accidental}\/ $U(1)_R$ 
symmetry, which nonetheless is broken by its coupling to messengers. 
Indeed, we show a very simple class of models of this type.  The models 
do not have a fundamental singlet field, eliminating aesthetic and 
various fine-tuning problems in cosmology and preserving the hierarchy. 
The gauginos and scalars in the supersymmetric standard model sector 
obtain flavor universal masses by standard model gauge interactions 
through loops of the messengers.  Given the absence of $U(1)_R$, 
there is no problem in generating gaugino masses, and no dangerous 
$R$-axion arises.

An explicit model that realizes our general philosophy is a 
supersymmetric $SU(N_c)$ QCD with massive vector-like quarks $Q^i$ 
and $\bar{Q}^i$ ($i = 1,\cdots,N_f$).  In addition, we introduce 
massive messengers $f$ and $\bar{f}$ and write the most general 
superpotential consistent with the gauge symmetry.  This is the 
entire model.  The important terms in the superpotential are 
given by
\begin{equation}
  W_{\rm tree} = m_{ij} \bar{Q}^i Q^j 
    + \frac{\lambda_{ij}}{M_{\rm Pl}} \bar{Q}^i Q^j \bar{f} f 
    + M \bar{f} f,
\label{eq:W_tree}
\end{equation}
where $\lambda_{ij}$ are coupling constants~\footnote{Here, we took 
  the scale of higher dimension operators to be the reduced Planck 
  scale $M_{\rm Pl}$ just for the sake of presentation, but of course 
  it can be some other scales as well.}.  (The effects of other terms 
will be discussed later.)  For concreteness, we take the messengers 
$f, \bar{f}$ to be in ${\bf 5} + {\bf 5}^*$ representations of $SU(5)$ 
in which the standard model gauge group is embedded.

Intriligator, Seiberg, and Shih (ISS) pointed out that supersymmetric 
$SU(N_c)$ QCD in the free magnetic phase ($N_c + 1 \leq N_f < 
\frac{3}{2} N_c$) breaks supersymmetry on a metastable local minimum 
if the quark masses $m_{ij}$ are much smaller than the dynamical scale 
$\Lambda$~\cite{Intriligator:2006dd}.  Note that in the ISS model 
a $U(1)_R$ symmetry is broken only down to $Z_{2N_c}$ which prevents 
the gaugino masses.  In the present model, however, the coupling to 
the messengers breaks it down to $Z_2$, so that the model does not 
have any $R$ symmetry beyond $R$-parity.

For the sake of concreteness, we discuss the case without the magnetic 
gauge group $N_f = N_c + 1$ below, although any $N_c + 1 \leq N_f 
< \frac{3}{2} N_c$ works equally well.  At energies below the dynamical 
scale, the non-perturbative low-energy effective superpotential is 
described as~\cite{Seiberg:1994bz}
\begin{equation}
  W_{\rm dyn} = \frac{1}{\Lambda^{2N_f-3}} 
    \bigl( \bar{B}_i M^{ij} B_j - {\rm det}M^{ij} \bigr),
\label{eq:W_dyn}
\end{equation}
where $M^{ij} = \bar{Q}^i Q^j$, $B_i = \epsilon_{i i_1 \cdots i_{N_c}} 
Q^{i_1} \cdots Q^{i_{N_c}}/N_c!$ and $\bar{B}_i = \epsilon_{i i_1 \cdots 
i_{N_c}} \bar{Q}^{i_1} \cdots \bar{Q}^{i_{N_c}}/N_c!$ are meson, baryon 
and antibaryon chiral superfields, respectively.  In the following, 
we adopt the basis in which the quark mass matrix is diagonal, $m_{ij} 
= -m_i \delta_{ij}$, with $m_i$ real and positive.  We also assume that 
they are ordered as $ m_1 > m_2 > \cdots > m_{N_f} > 0$ without loss 
of generality.  Here, we have taken all masses different to avoid 
(potentially) unwanted Nambu--Goldstone bosons.

In terms of fields with canonical dimensions $S^{ij} = M^{ij}/\Lambda$, 
$b_i = B_i/\Lambda^{N_f-2}$ and $\bar{b}_i = \bar{B}_i/\Lambda^{N_f-2}$, 
the dynamical superpotential of Eq.~(\ref{eq:W_dyn}) together with the 
quark mass terms (the first term of Eq.~(\ref{eq:W_tree})) can be written 
as~\footnote{The fields $S^{ij}$, $b_i$ and $\bar{b}_i$ are in general 
  not canonically normalized by incalculable $O(1)$ wavefunction 
  renormalization factors, which are not important to our discussions 
  and hence disregarded in the rest of the letter.}
\begin{equation}
  W_{\rm ISS} = \bar{b}_i S^{ij} b_j 
    - \frac{{\rm det} S^{ij}}{\Lambda^{N_f-3}} - m_i \Lambda S^{ii}.
\label{eq:W_ISS}
\end{equation}
For $N_f > 3$, the superpotential term ${\rm det} S^{ij}$ is irrelevant 
and can be ignored to discuss physics around the origin $S^{ij} 
= 0$~\footnote{This term, however, is important to see that there are 
  global supersymmetric minima at nonzero $S^{ij}$ as suggested by 
  the general arguments.}.  The superpotential of Eq.~(\ref{eq:W_ISS}) 
then leads to a local minimum at
\begin{equation}
  b = \bar{b} 
  = \left( \begin{array}{c} 
      \sqrt{m_1 \Lambda} \\ 0 \\ \vdots \\ 0 
    \end{array} \right),
\qquad
  S^{ij} = 0,
\label{eq:local-min}
\end{equation}
where supersymmetry is broken because $F_{S^{ij}} = -(\partial_{S^{ij}}W)^* 
= m_i \delta_{ij} \Lambda \neq 0$ for $i,j \neq 1$.  Even though $S^{ij}$ 
$(i, j \neq 1)$ are classically flat directions, they are lifted by the 
one-loop Coleman--Weinberg potential.  As a result, the origin $S^{ij}=0$ 
is a local minimum, with curvature $m_{S^{ij}}^2 \sim m \Lambda/16\pi^2$ 
for all $m_i \sim m$.  It is long-lived as long as $m_i \ll \Lambda$, 
where the weakly-coupled analysis of the low-energy theory is valid.

The existence of a supersymmetry breaking minimum of Eq.~(\ref{eq:local-min}) 
can be viewed as a result of an accidental (and approximate) $U(1)_R$ 
symmetry possessed by the superpotential of Eq.~(\ref{eq:W_ISS}) with 
the $R$-charge assignments $R(S^{ij})=2$, $R(b_i)=R(\bar{b}_i)=0$, in the 
limit of neglecting the irrelevant term of ${\rm det} S^{ij}/\Lambda^{N_f-3}$. 
In fact, this accidental $U(1)_R$ symmetry is also a reason for the 
origin $S^{ij}=0$ being the minimum of the effective potential as 
a symmetry enhanced point.  This picture is corrected by the coupling 
of $Q^i$ and $\bar{Q}^i$ to the messengers and by higher dimension terms 
in the superpotential omitted in Eq.~(\ref{eq:W_tree}), which introduce 
$U(1)_R$ violating effects to the supersymmetry breaking sector.  These 
effects, however, can be easily suppressed as we will see later, and 
the basic picture described above can be a good approximation of the 
dynamics.

At the supersymmetry breaking minimum of Eq.~(\ref{eq:local-min}) 
(with $S^{ij}$ slightly shifted due to $U(1)_R$ violating effects), 
the messenger fields have both supersymmetric and holomorphic 
supersymmetry breaking masses:
\begin{equation}
  M_{\rm mess} 
  = M + \frac{\lambda_{ij} \Lambda}{M_{\rm Pl}} \langle S^{ij} \rangle 
  \simeq M,
\label{eq:M_mess}
\end{equation}
and
\begin{equation}
  F_{\rm mess} 
  = \frac{\lambda_{ij} \Lambda}{M_{\rm Pl}} F_{S^{ij}} 
  = \frac{\bar{m} \Lambda^2}{M_{\rm Pl}},
\label{eq:F_mess}
\end{equation}
where
\begin{equation}
  \bar{m} \equiv \sum_{i\neq 1} \lambda_{ii} m_i.
\label{eq:m-bar}
\end{equation}
The usual loop diagrams of the messenger fields then induce gauge-mediated 
scalar and gaugino masses in the supersymmetric standard model sector, 
of the magnitude~\cite{Dine:1981gu,Dine:1994vc}
\begin{equation}
  m_{\rm SUSY} 
  \simeq \frac{g^2}{16\pi^2} \frac{\bar{m} \Lambda^2}{M M_{\rm Pl}},
\label{eq:m_SUSY}
\end{equation}
where $g$ represents generic standard model gauge coupling constants.

Several conditions for the parameters need to be met for the model to 
be phenomenologically successful.  Even though not necessary, we regard 
all the quark masses (and the couplings $\lambda_{ij}$) to be comparable, 
$m_i \sim m$ ($\lambda_{ij} \sim \lambda$), in the numerical estimates 
below.

First, we would like $m_{\rm SUSY}$ to stabilize the electroweak scale, 
and hence $m_{\rm SUSY} = O(100~{\rm GeV}\!\sim\!1~{\rm TeV})$.  This 
corresponds to
\begin{equation}
  \frac{\bar{m} \Lambda^2}{M M_{\rm Pl}} \approx 100~{\rm TeV}.
\label{eq:naturalness}
\end{equation}
On the other hand, we would like the gauge-mediated contribution to 
the scalar masses dominate over the gravity-mediated piece to avoid 
excessive flavor-changing processes, leading to $m_{3/2} \approx 
m \Lambda/M_{\rm Pl} \lesssim 10^{-2} m_{\rm SUSY}$.  Therefore,
\begin{equation}
  m M \lesssim 10^{-4} \bar{m} \Lambda.
\label{eq:gravity}
\end{equation}
We also need the messengers to be non-tachyonic,
\begin{equation}
  M^2 > \frac{\bar{m} \Lambda^2}{M_{\rm Pl}}.
\label{eq:tachyon}
\end{equation}
In addition, the analysis of supersymmetry breaking is valid only if 
$m$ is sufficiently smaller than $\Lambda$:
\begin{equation}
  m \lesssim 0.1 \Lambda.
\label{eq:validity}
\end{equation}

We now discuss the effects of $U(1)_R$ violation.  These effects 
cause shifts of $S^{ij}$ from the origin, which must be smaller than 
$\approx 4\pi \sqrt{m \Lambda}$ for the ISS analysis to be valid, and 
than $\approx M M_{\rm Pl}/\lambda \Lambda$ to avoid tachyonic messengers. 
One origin of $U(1)_R$ violation comes from higher dimension terms in 
the superpotential, omitted in Eq.~(\ref{eq:W_tree}).  The dominant 
effect comes from
\begin{equation}
  \Delta W = \frac{\lambda_{ijkl}}{M_{\rm Pl}} \bar{Q}^i Q^j \bar{Q}^k Q^l 
  = \frac{\lambda_{ijkl} \Lambda^2}{M_{\rm Pl}} S^{ij} S^{kl}.
\label{eq:SS}
\end{equation}
These terms may destabilize the minimum, since they lead to linear 
terms of $S^{ij}$ in the potential through $F_{S^{ij}} = m_i \delta_{ij} 
\Lambda$~\cite{Kitano:2006xg}.  The squared masses of $S^{ij}$ from the 
one-loop effective potential are $m_{S^{ij}}^2 \sim m \Lambda/16\pi^2$, 
while the linear terms are $\sim (\lambda_{ijkk} m_k \Lambda^3/M_{\rm Pl}) 
S^{ij}$.  Therefore, the shifts of the fields are $\Delta S^{ij} \sim 
16\pi^2 \lambda_{ijkk} \Lambda^2/M_{\rm Pl}$.  Requiring this to be 
sufficiently small, we obtain the condition
\begin{equation}
  \frac{\lambda_{ijkk} \Lambda^2}{M_{\rm Pl}} 
  \lesssim {\rm min} \biggl\{ 0.1 (m \Lambda)^{1/2}, 
    10^{-2} \frac{M M_{\rm Pl}}{\lambda \Lambda} \biggr\}.
\label{eq:stability}
\end{equation}
Similar conditions can be worked out for even higher order terms, but 
they are rather mild.

Another source of $U(1)_R$ violation comes from the coupling of $Q^i$ 
and $\bar{Q}^i$ to the messengers, which shifts the minimum of $S^{ij}$ 
at the loop level.  The effect of the messengers on the $S^{ij}$ effective 
potential can be calculated by computing the one-loop Coleman--Weinberg 
potential arising from the last two terms of Eq.~(\ref{eq:W_tree}):
\begin{equation}
  W_{\rm mess} = \frac{\lambda_{ij} \Lambda}{M_{\rm Pl}} S^{ij} \bar{f} f 
    + M \bar{f} f.
\label{eq:W_mess}
\end{equation}
The resulting effective potential takes the following generic form
\begin{equation}
  \Delta V \approx \frac{\bar{m}^2 \Lambda^4}{16 \pi^2 M_{\rm Pl}^2} 
  {\cal F}\left(\frac{\lambda_{ij} \Lambda S^{ij}}{M M_{\rm Pl}}\right),
\label{eq:Delta-V}
\end{equation}
where ${\cal F}(x)$ is a real polynomial function with the coefficients 
of $O(1)$ up to symmetry factors.  The resulting shifts of $S^{ij}$ are 
of order $\lambda^3 m \Lambda^4/M M_{\rm Pl}^3$, which are sufficiently 
small if
\begin{equation}
  M \gtrsim \frac{\lambda^2 m^{1/2} \Lambda^{5/2}}{M_{\rm Pl}^2}.
\label{eq:cond-R}
\end{equation}
Note that the coupling to the messengers in Eq.~(\ref{eq:W_mess}) does 
not generate a new supersymmetric minimum.  However, turning on the 
expectation values for the messengers may allow for lowering the vacuum 
energy, depending on the combinations of $m_{ij}$ and $\lambda_{ij} 
\bar{f} f$.  Even if this is the case, the tunneling to a lower 
minimum at $\bar{f} f \approx m M_{\rm Pl}/\lambda$ can easily be 
made suppressed to the level consistent with the longevity of our 
universe, if $M M_{\rm Pl}/\lambda \gtrsim m^{1/2} \Lambda^{3/2}$.

It is now easy to see that there is a wide range of parameters that 
satisfy the conditions Eqs.~(\ref{eq:naturalness},~\ref{eq:gravity},%
~\ref{eq:tachyon},~\ref{eq:validity},~\ref{eq:stability},~\ref{eq:cond-R}). 
For instance, if we take $\lambda_{ij} \sim \lambda_{ijkl} \sim 1$, 
$\Lambda \sim 10^{11}~{\rm GeV}$, $m \sim \bar{m} \sim 10^{8}~{\rm GeV}$ 
and $M \sim 10^{7}~{\rm GeV}$, then all the requirements are easily 
satisfied.  Note that the conditions of Eqs.~(\ref{eq:stability},%
~\ref{eq:cond-R}) are generically rather weak, unless $\Lambda$ is 
close to $M_{\rm Pl}$.  This is because the relevant interactions 
in Eqs.~(\ref{eq:SS},~\ref{eq:W_mess}) arise from higher dimension 
operators suppressed by $M_{\rm Pl}$.

Finally, we discuss if there are any unwanted light fields in the 
model.  The fermionic fields in $S^{ij}$ $(i,j \neq 1)$ are massless 
in the ISS model, but they acquire masses here due to the generic terms 
in Eq.~(\ref{eq:SS})~\footnote{Of course, one of these fields remains 
  massless as the Goldstino which is eaten by the gravitino.}.  They 
can decay to standard model particles through their coupling to the 
messengers and hence harmless.  There is a Nambu--Goldstone boson (NGB) 
of a spontaneously broken $U(1)_B$ symmetry, $b^1-\bar{b}^1$, and its 
fermionic partner.  Exactly massless NGB and fermion would be a radiation 
component of the universe.  Their abundance is diluted by an order of 
magnitude due to the QCD phase transition and is in general consistent 
with the constraint from the big-bang nucleosynthesis, $\Delta N_\nu 
\lesssim 1.5$~\cite{Cyburt:2004yc}.  Alternatively, they can be made 
massive by gauging $U(1)_B$, or avoided entirely by employing an 
$SO(N_c)$ or $Sp(N_c)$ gauge group for supersymmetry breaking, instead 
of $SU(N_c)$.  The gravitino is the lightest supersymmetric particle 
and hence stable if $R$-parity is unbroken.  It places an upper limit 
on the reheating temperature~\cite{Moroi:1993mb}, which is acceptable 
{\it e.g.}\/, in leptogenesis models by non-thermal production of 
right-handed scalar neutrinos~\cite{Murayama:1992ua}.

In summary, we advocated gauge mediation models of supersymmetry 
breaking with generic superpotentials without $U(1)_R$ symmetry.  Using 
metastable minima, we find a class of phenomenologically successful 
models without any elementary gauge singlet fields.  We find the 
simplicity and generality of the models quite remarkable.

\begin{acknowledgments}
  This work was supported in part by the U.S. DOE under Contract
  DE-AC03-76SF00098, and in part by the NSF under grant PHY-04-57315.
  The work of Y.N. was also supported by the NSF under grant
  PHY-0555661, by a DOE OJI award, and by an Alfred P. Sloan Research
  Fellowship.
\end{acknowledgments}

\end{document}